\documentclass[aps,prl,superscriptaddress,twocolumn,showpacs,preprintnumbers,amsmath,amssymb]{revtex4-1}
\usepackage{graphicx}
\usepackage{dcolumn}
\usepackage{bm}

\begin{document}

\title{Radiative annihilation of a soliton and an antisoliton in the coupled sine-Gordon equation}

\author{V. M. Krasnov}
\email[E-mail:]{Vladimir.Krasnov@fysik.su.se}
\affiliation{Department of Physics, Stockholm University, AlbaNova
University Center, SE-10691 Stockholm, Sweden }
\date{\today }

\begin{abstract}

In the sine-Gordon equation solitons and antisolitons in the
absence of perturbations do not annihilate. Here I present
numerical analysis of soliton-antisoliton collisions in the
coupled sine-Gordon equation. It is shown that in such a system
soliton-antisoliton pairs (breathers) do annihilate even in the
absence of perturbations. The annihilation occurs via a
logarithmic-in-time decay of a breather caused by emission of
plasma waves in every period of breather oscillations. This also
leads to a significant coupling between breathers and propagating
waves, which may lead to self-oscillations at the geometrical
resonance conditions in a dc-driven system. The phenomenon may be
useful for achieving superradiant emission from coupled
oscillators.


\end{abstract}

\maketitle

\section{Introduction}

Analysis of soliton dynamics in the sine-Gordon (SG) formalism is
important in many research areas \cite{Kivsh_Malomed_RMP},
including non-linear optics, condensed matter physics, atomic
\cite{Braun} and particle physics \cite{QuantumSG}. Solitons are
elementary particles of the sine-Gordon equation, in a sense that
they are quantized and do not spontaneously decay.
Soliton-antisoliton collision is a nontrivial example of
interaction of strongly non-linear waves. It may lead either to
passage of the two waves or to formation of a bound pair - the
breather \cite{Scott}. Within the pure SG equation soliton and
antisoliton do not annihilate because the collision is elastic and
the annihilation is prohibited by energy conservation. However,
addition of various perturbation terms to the SG equation does
allow particle-antiparticle annihilation via breather decay
\cite{Scott,PedersenAnnih}. This may happen both via intrinsic
viscous damping and via radiative losses from the breather
\cite{Kivsh_Malomed_RMP}.

In recent years properties of solitons in the coupled sine-Gordon
equation (CSGE) are being actively studied. The CSGE describes
complex behavior of interacting systems, such as atoms in a
periodic potential \cite{Braun}, magnetic multilayers
\cite{MagneticML}, stacked Josephson junctions
\cite{SBP,SakUstFiske,Modes,Fluxon} and layered superconductors
\cite{KleinerMuller94,KatterweFrau,KatterweFiske}. Coupling of $N$
systems leads to a variety of unusual effects. First of all, it
leads to appearance of $N$ eigenmodes with different symmetries,
length scales and velocities \cite{KleinerModes,SakUstFiske}. Even
though the exact soliton solution in this case is not known,
numerical and approximate analytic results demonstrated that the
soliton becomes composed of different eigenmodes
\cite{Modes,Fluxon} and the shape of such a composite soliton may
become very unusual in the dynamic case. Next, unlike the
sine-Gordon equation, the coupled sine-Gordon equation is not
Lorentz-invariant \cite{KatterweFiske}. Therefore superluminal
soliton motion (faster than the slowest eigenmode velocity) is
possible \cite{Modes,Cherenkov,KatterweFiske}. It is accompanied
by Cherenkov-type radiation, due to decomposition of soliton
components with eigenmode velocities slower than the speed of the
soliton into plasma waves travelling along with the soliton
\cite{Modes,Fluxon,Cherenkov}.

In this work I present numerical analysis of soliton-antisoliton
collisions within the coupled sine-Gordon equation with focus on
Josephson vortex (fluxon) dynamics in magnetically coupled stacked
Josephson junctions. Both direct (fluxon and antifluxon in the
same junction) and indirect (fluxon and antifluxon in different
junctions) collisions are considered. It is demonstrated that
soliton-antisoliton pair in the CSGE can annihilate even in the
absence of viscous damping or other perturbations. Annihilation
occurs via emission of plasma waves from an oscillating breather,
to some extent resembling annihilation of elementary particles via
emission of a pair of photons. The radiative annihilation leads to
a significant coupling of a breather to linear waves and brings
about a variety of resonant and self-oscillation phenomena
\cite{Breather}, which can be useful for achieving a coherent
superradiant emission from coupled systems
\cite{Ozyuzer,Breather}.

\section{General relations}

We consider one-dimensional chains/junctions described by the
perturbed sine-Gordon equation:
\begin{equation}
\varphi^{\prime \prime }-\ddot \varphi -\alpha\dot \varphi = \sin
\varphi - \gamma, \label{Pert_SG}
\end{equation}
where $\varphi$ is the phase variables, ``primes'' and ``dots''
denote spatial $\varphi^\prime=\partial\varphi/\partial x$ and
temporal $\dot \varphi=\partial\varphi/\partial t$ derivatives,
$\alpha$ is the viscous damping parameter and $\gamma$ is the
driving (bias) term. In the absence of perturbation terms
$\alpha=\gamma=0$ it reduces to the pure SG equation:
\begin{equation}
\varphi^{\prime \prime }-\ddot \varphi = \sin \varphi. \label{SG}
\end{equation}

The soliton in the SG Eq.(\ref{SG}) is a $2\pi$ phase kink
\cite{Scott}:
\begin{equation}\label{Soliton}
F = 4 \arctan [\exp (x-ut)/\sqrt{1-u^2}],
\end{equation}
where $u$ is the velocity of the soliton, normalized by the speed
of light (the Swihart velocity). The velocity dependent factor
represents the relativistic contraction of the soliton when its
velocity approaches the speed of light $u \rightarrow 1$
\cite{Scott}. This is the consequence of Lorentz invariance of the
SG equation (\ref{SG}). The normalized energy of a static soliton
$u=0$ is $E_{sol}=8$.

\subsection{The coupled sine-Gordon equation}

We assume that a system of $N$ interacting junctions can be
described by the perturbed CSGE \cite{SBP}:
\begin{equation}
\varphi_i^{\prime \prime }={\bf A_{ij}}\left[\ddot \varphi_j
+\alpha\dot \varphi _j + \sin \varphi _j - \gamma \right].
\label{CSGE}
\end{equation}
Here $i,j=1,2, ... N$ is the junction index and ${\bf A}_{ij}$ is
the coupling matrix, off-diagonal elements of which describe
interaction between different junctions. In what follows we will
consider the simplest case of nearest neighbor interaction,
described by a symmetric tridiagonal matrix the only nonzero
elements of which are:

$A_{i,i}=1,~ A_{i,i-1}= A_{i,i+1}=-S$.

\noindent Here $S<0.5$ is the coupling strength. The CSGE can be
also written in the equivalent inverted form
\begin{equation}
{\bf A_{ij}^{-1}}\varphi_i^{\prime \prime }-\ddot \varphi_j
-\alpha\dot \varphi_j = \sin \varphi _j - \gamma. \label{CSGE_Inv}
\end{equation}
Apparently, for $N=1$ it reduces to the perturbed SG Eq.
(\ref{Pert_SG}). In case of two coupled chains $N=2$, the system
of unperturbed CSGE $\alpha=\gamma=0$ reads:
\begin{eqnarray}
\frac{1}{1-S^2}\varphi_1^{\prime \prime }-\ddot \varphi_1 = \sin
\varphi_1 - \frac{S}{1-S^2}\varphi_2^{\prime
\prime },\\
\frac{2}{1-S^2}\varphi_1^{\prime \prime }-\ddot \varphi_2 = \sin
\varphi_2 - \frac{S}{1-S^2}\varphi_1^{\prime \prime }.
\label{DoubleCSGE}
\end{eqnarray}
It is easy to verify by direct application of the Lorentz
transformation that the CSGE is not Lorentz invariant, unlike the
SG equation (\ref{SG}).

Physically, the considered type of coupling corresponds e.g., to
magnetic (inductive) interaction of stacked Josephson junctions
\cite{SBP}. In this case space and time in the dimensionless
equations are normalized by the Josephson penetration depth,
$\lambda _{J}$, and the Josephson plasma frequency $\omega _{p}$,
respectively, the velocity is normalized by the Swihart velocity
$c_0=\lambda_J \omega_p$ and $\gamma$ by the Josephson critical
current $\gamma=I/I_c$. More details on the normalization and the
formalism can be found in Refs.\cite{Modes,Fluxon}. As mentioned
in the introduction, coupling terms as in Eqs.
(\ref{CSGE},\ref{CSGE_Inv}) are also relevant for other objects,
like atomic chains \cite{Braun} and magnetic multilayers
\cite{MagneticML}.

The energy density of the coupled system is \cite{Modes,Fluxon}
\begin{equation}
\frac{\partial E(x)}{\partial x} =\frac{1}{2}\varphi_j^\prime{\bf
A_{ij}^{-1}}\varphi_i^\prime + \sum_{i=1}^{N}(1-\cos \varphi_i)+
\frac{1}{2}\dot\varphi_i^2. \label{Energy}
\end{equation}
Here the first, the second and the third terms represent
correspondingly the magnetic/elastic, the Josephson/potential and
the electric/kinetic energies for the case of a junction/chain.

Coupling leads to splitting of the dispersion relation of small
oscillations into $N$ eigenmodes with different symmetries and
propagation velocities \cite{KleinerModes,SakUstFiske}:

\begin{equation}\label{c_n}
c_n=\left[ 1-2S\cos\frac{\pi n}{N+1} \right]^{-1/2},
~(n=1,2,...,N).
\end{equation}
The slowest mode $n=N$ corresponds to out-of-phase (antisymmetric)
oscillations in neighbor junctions and the fastest, $n=1$ to the
in-phase (symmetric) oscillations in all the junctions
\cite{Note1}.

\subsection{A single soliton in the unperturbed CSGE}

For the solitonic motion with a constant velocity $u$,
$\varphi(x,t)=\varphi(x-ut)$, the unperturbed CSGE Eq.
(\ref{CSGE_Inv}) with $\alpha=\gamma=0$ can be written in the
simple vector form \cite{Fluxon}:
\begin{equation}\label{VectorCSGE}
[{\bf A_{ij}^{-1}} -u^2 {\bf E}]{\bf \varphi}^{\prime \prime} =
\sin{\bf \varphi}.
\end{equation}
Here ${\bf E}$ is the unitary matrix. This equation is essentially
similar to the static CSGE, for which the first integral is known
\cite{Modes}. Therefore we can in a similar manner write the first
integral for the solitonic motion:
\begin{equation}\label{FirstInt}
\frac{1}{2} \varphi_j^{\prime} [{\bf A_{ij}^{-1}} -u^2 {\bf E}]
\varphi_i^{\prime} - \Sigma_{i=1}^N [1-\cos\varphi_i] = C.
\end{equation}
For a single soliton the constant $C=0$ because at the infinity
$\varphi_i=\varphi_i^{\prime}=0$. From comparison with the general
expression for the energy density Eq. (\ref{Energy}), it is easily
seen that the soliton energy is twice the magnetic/elastic energy
\begin{equation}\label{SolitEnergy}
E_{sol}=2E_m=\int dx \varphi_j^\prime{\bf
A_{ij}^{-1}}\varphi_i^\prime,
\end{equation}
as is also the case for the soliton in the SG equation
\cite{Scott}.

The exact soliton solution in the CSGE is not yet known. However,
an approximate composite soliton solution has been proposed,
verified by numerical simulations and by perturbation correction
calculations \cite{Modes,Fluxon}. It is represented by a linear
superposition of solitonic waves Eq. (\ref{Soliton}),
corresponding to different eigenmodes.
\begin{eqnarray}\label{CompSolit}
\varphi_i \simeq \sum_{n=1}^N \kappa_{n,i} F_n,~(i=1,2,...,N),\\
F_n = 4 \arctan \left[\exp
\left(\frac{x-ut}{\lambda_n\sqrt{1-u^2/c_n^2}}\right)\right].
\end{eqnarray}
Here $i$ is the junction number, $\lambda_n$ is the characteristic
length scale of the eigenmode $n$:
\begin{equation}\label{Lambda_n}
\lambda_n=\left[ 1-2S\cos\frac{\pi n}{N+1} \right]^{-1/2},
~(n=1,2,...,N).
\end{equation}
Note that $\lambda_n^{-2}$ are eigenvalues of the coupling matrix
${\bf A_{ij}}$ and coefficients $\kappa_{n,i}$ are components of
the eigenvectors of ${\bf A_{ij}}$, normalized so that
$\sum_{n=1}^N \kappa_{n,i} = 1$ in the junction containing the
soliton and zero in all other junctions \cite{Fluxon,Note1}. Thus
the soliton consists of a $2\pi$ kink in one junction and ripples
in all other junctions. The soliton shape (coefficients
$\kappa_{n,i}$) does depend on the junctions number and is, for
example, different for the soliton in the outmost and in the
central junctions of the stack. Amplitudes of ripples in the
neighbor junctions depend on the coupling strength and can be
significant in the strong coupling case $S\simeq 0.5$. The ripples
decrease with the distance from the soliton both along and across
the junctions.

As discussed in Ref. \cite{Modes}, the static soliton energy in
the CSGE is larger than that in the single SG equation both due to
presence of ripples in neighbor chains and reconstruction of
characteristic length scales Eq.(\ref{Lambda_n}). Let's, for
example, estimate the energy in the simplest case $N=2$. In this
case the multicomponent soliton, Eq. (\ref{CompSolit}), becomes
\cite{Note1}
\begin{equation}\label{Esol_N2}
N=2:\left\{
\begin{array}{c}
\varphi_1=\frac{F_1+F_2}2, ~\lambda_1=\left( 1-S\right) ^{-1/2}\\
\varphi_2=\frac{F_1-F_2}2, ~\lambda_2=\left( 1+S\right) ^{-1/2}
\end{array}
\right.
\end{equation}
Substituting those into Eq. (\ref{SolitEnergy}) and taking into
account that $A_{1,1}^{-1}=A_{2,2}^{-1}=1/(1-S^2)$ and
$A_{1,2}^{-1}=A_{2,1}^{-1}=S/(1-S^2)$ we obtain
\begin{equation}\label{Esol_N2}
\begin{array}{c}
E_{sol}(N=2)=\frac{2}{1-S^2}\left[
\lambda_1^{-1}(1+S)+\lambda_2^{-1}(1-S)\right] = \\
4[(1+S)^{-1/2}+(1-S)^{-1/2}].
\end{array}
\end{equation}
For $S=0.5$ we get $E_{sol}(N=2)\simeq 8.92$, which is $\simeq
1.12$ times larger than $E_{sol}(N=1)=8$ for a single junction, in
good agreement with numerical simulations \cite{Modes}. The
soliton energy increases with $N$ and saturates at $\simeq 3.6
E_{sol}(N=1)$ in the strong coupling case $S=0.5$
\cite{CommentFl}.

The shape of the soliton in the CSGE experiences strong
metamorphoses in the dynamic case \cite{Modes,Fluxon}. Indeed,
since the soliton components $F_n$, Eq. (\ref{CompSolit}),
experience Lorentz contraction at different characteristic
velocities $c_n$, Eq. (\ref{c_n}), the relative shape of the
soliton does not remain the same as in the static case. When $u$
approaches the slowest velocity $c_N$, the corresponding component
$n=N$ gets contracted, while the rest of the soliton remains
uncontracted. Such the partial Lorentz contraction was confirmed
by numerical simulations \cite{Modes,Fluxon}. The soliton survives
even at superluminal velocity $u>c_N$, however in this case the
characteristic length $\lambda_N$ becomes imaginary due to the
Lorentz factor. This corresponds to transformation of the
corresponding soliton component $F_N$ into the out-of-phase plasma
wave \cite{Modes}. A similar process of decomposition of soliton
components into plasma waves with corresponding symmetries occurs
when $u$ exceeds any of the characteristic velocities $c_n$
\cite{Fluxon}. The phenomenon resembles Cherenkov emission from a
superluminal particle \cite{Modes,Cherenkov,Fluxon} with the
exception that the speed of light is multiple-valued, Eq.
(\ref{c_n}), and the dispersion relation is different.

Multisoliton states in the CSGE are dominated by a profound
metastability \cite{Modes,Compar,KatterweFrau}, i.e., for given
boundary conditions a large variety of metastable soliton
distributions is possible. Moving solitons interact with linear
waves, which leads to appearance of geometrical resonances
(standing waves) in finite size systems. Note that a soliton in
the CSGE can excite all eigenmodes Eq.(\ref{c_n}), which leads to
a large variety of geometrical resonances
\cite{SakUstFiske,KatterweFiske}.

\begin{figure*}
\includegraphics[width=6.5in]{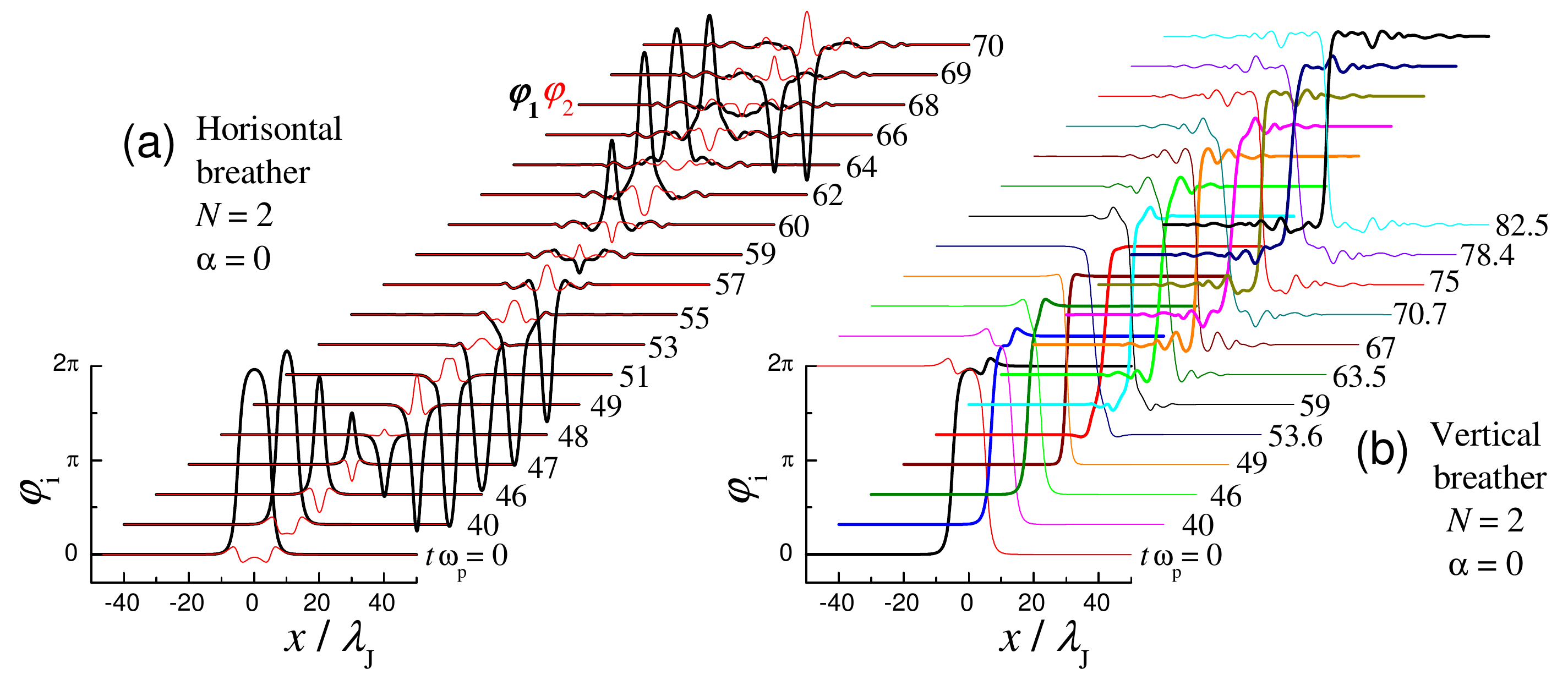}
\label{Fig1}
\caption{(Color online). Snapshots of time evolution of phase
distributions upon soliton-antisoliton collision in the
unperturbed ($\alpha = \gamma = 0$) double-junction CSGE for (a)
direct collision of a solton and an antisoliton in the junction-1
and (b) indirect collision of a soliton in the junction-1 and an
antisoliton in the junction-2. Thick and thin lines represent
$\varphi_1$ and $\varphi_2$, respectively. It is seen that both
the horizontal (a) and vertical (b) breathers are decaying due to
emission of plasma waves, even in the absence of perturbations. }
\end{figure*}

\subsection{Breathers}

Breather is a bound soliton-antisoliton pair. The breather
solution of the SG equation Eq. (\ref{SG}) in the center-of-mass
frame is \cite{Scott}
\begin{equation}\label{Breather}
\varphi = 4 \arctan \left(\frac{\tan \nu \sin[(\cos \nu)
t]}{\cosh[(\sin \nu)x]} \right).
\end{equation}
Here $0<\nu<\pi/2$ is determining the breather amplitude
$\varphi_{Br}=4 \nu$. The breather is oscillating without
annihilation or decay at a frequency $\omega_{Br} = \cos \nu$,
which is always less than the plasma frequency $\omega_{Br}<1$.
The solution Eq. (\ref{Breather}) is valid for an infinite system
$L=\infty$. A more complicated solution for the finite size system
can be found in Ref. \cite{Costabile}. The total energy of the
breather $E_{Br} = 16 \sin \nu$ is smaller than the energy of two
static solitons $E_{Br}<2E_{sol}=16$, leading to binding of the
soliton and the antisoliton.

Finite dissipation $\alpha >0$ leads to decay of the breather and
facilitates soliton-antisoliton annihilation. The decay is
primarily caused by the viscous damping of the soliton-antisoliton
motion. However, minor radiative losses also appear
\cite{Kivsh_Malomed_RMP}. A qualitative change of the wave form
takes place upon the soliton-antisoliton annihilation. Initially,
the soliton-antisoliton pair ($4\nu \simeq 2 \pi$) shrinks, i.e.,
the maximum separation between the pair $\Delta x \sim 1/\tan \nu
\gg 1$ gradually decreases after every collision. At $4\nu
\lesssim \pi$, the soliton and the antisoliton completely merge
and can no longer be distinguished. Further decay (reduction of
$\nu$) leads to expansion of the breather. From Eq.
(\ref{Breather}) it follows that for small $\nu \ll 1$ the size of
the breather is $\propto 1/\sin \nu$. Eventually, the breather
turns into the longitudinal plasma wave with $\omega_{Br}=\cos \nu
\simeq 1$ and the wave number $k_x \simeq \sin \nu \simeq 0$. This
accomplishes the soliton-antisoliton annihilation.

Breathers play role not only in soliton-antisoliton annihilation
and the opposite (time-reversal) process of creation (or
penetration) of the soliton \cite{CommentFl}. Breathers also
interact with traveling waves and external forces
\cite{Kivsh_Malomed_RMP}. In Ref. \cite{Breather} it was argued
that breathers in the CSGE can help to pump energy from the
external dc-power supply ($\gamma$ term in Eq. (\ref{CSGE})) into
the oscillating travelling waves, which leads to appearance of
self-oscillations in the dc-driven CSGE with $\gamma>0$. The
phenomenon may find practical applications for generation of
coherent (superradiant) THz sources based on stacked intrinsic
Josephson junctions in high-temperature superconductors
\cite{Ozyuzer}.

In the CSGE we will consider two distinctly different types of
breathers (referred to as the horizonal and the vertical):

(i) the ''horizontal" breather corresponds to a direct collision
of a soliton and an antisoliton in the same junction.

(ii) the ''vertical" breather corresponds to an indirect collision
of a soliton and an antisoliton in different junctions. Unlike the
horizontal breather, the vertical breather does not annihilate
even in the presence of dissipation, but leads to formation of a
stable static soliton-antisoliton pair
\cite{Compar,KleinerAntiFlux}. For $N=2$ the static ''vertical"
soliton-antisoliton pair has an exact antisymmetric solution
\cite{SBP}: $\varphi_1=-\varphi_2 = F_2$. The energy of the
vertical pair
\begin{equation}\label{E_vertical2}
E_{vert}(N=2)=2\times 8\lambda_2^{-1} = 16(1+S)^{-1/2}
\end{equation}
is smaller than twice the isolated soliton energy
Eq.(\ref{Esol_N2}), leading to binding of the pair.

\subsection{Numerical procedure}

The system of partial differential equations Eq.(\ref{CSGE_Inv})
for different $N$ and junction length $L$ is solved numerically
using an explicit finite difference method (central difference in
space and time). The spatial mesh size $\Delta x$ was typically
0.025 and the temporal $\Delta t = \Delta x/10$. The absence of
spurious effects was checked by changing mesh sizes and
integration times.

Static solitons and antisolitons Eq. (\ref{Soliton}) were
introduced at certain positions at the initial time. The system is
then given long enough time to relax with a large damping factor
$\alpha = 2$. The large viscosity prevents significant soliton
motion during the transient period. After that calculations
continued with the desired value of $\alpha$. The time count $t=0$
starts from the end of the transient period.

All simulations were made for zero external field boundary
conditions at $x=\pm L/2$
\begin{equation}\label{BoundaryCond}
\partial \varphi_i / \partial x  = 0.
\end{equation}
Those boundary conditions are non-radiative, i.e., preclude energy
flow through the edges \cite{Hu}. In some cases dynamic radiative
boundary conditions were employed (still at zero external field)
following Ref. \cite{TheoryFiske}. The radiation emission is
facilitated by the finite radiation impedance $Z$. For more
details see Ref. \cite{TheoryFiske}.

All presented simulations are done for the strong coupling case
$S\simeq 0.5$, close to the maximum value, relevant e.g. for
atomic scale intrinsic Josephson junctions
\cite{KleinerMuller94,Fluxon,KatterweFiske,KatterweFrau}. It was
checked that variation of the coupling strength does not affect
the qualitative presence of the effects described below.

\begin{figure}[t]
\includegraphics[width=2.7 in]{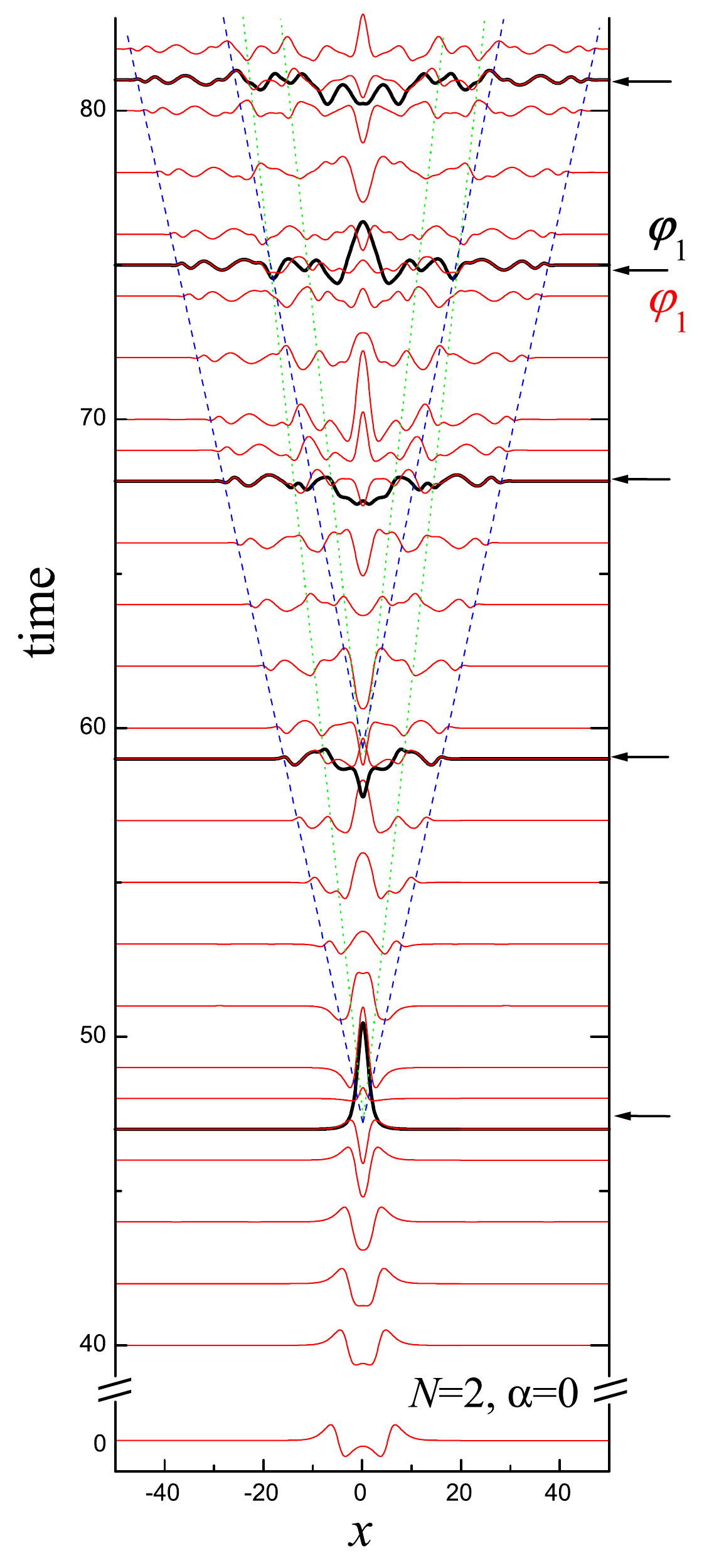}
\label{Fig2} \caption{ (Color online). Detailed view of the time
evolution of $\varphi_2$ (thin lines) for the horizontal breather
from Fig. 1 (a). The phase $\varphi_1$ (thick lines) is shown
close to the moments of collisions, marked by arrows. It is seen
that emission of two wave fronts occurs at every collision: the
fast front (marked by blue dashed lines) has an in-phase symmetry
$\varphi_1=\varphi_2$ and propagates with the fast velocity $c_1$,
the slow front (marked by green dotted lines) has an out-of-phase
symmetry $\varphi_1=-\varphi_2$ and propagates with the slow
velocity $c_N$.}
\end{figure}

\section{Results}

\subsection{Unperturbed soliton-antisoliton dynamics}

Figure 1 shows time sequence of calculated phase profiles
$\varphi_1$ (thick lines) and $\varphi_2$ (thin lines) for the
unperturbed ($\alpha = \gamma = 0$) CSGE in a double junction
system $N=2$, Eq. (\ref{DoubleCSGE}), for the horizontal (a) and
the vertical (b) breathers. Initially at $t=0$ the soliton and the
antisoliton are well separated. The solitons collide for the first
time approximately at the same time ($t_1\simeq 47$). For a single
SG equation, the breather Eq. (\ref{Breather}) would continue to
oscillate without decay with the same periodicity, i.e. the
subsequent collision would occur at time intervals $2t_1$. This is
clearly not the case in the unperturbed CSGE:

(i) First of all, subsequent collisions occur at smaller time
intervals. For example, the second collision for both breather
occurs at $t_2\simeq 59$ and $t_2-t_1 \simeq 12$ much shorter than
$2t_1 \simeq 94$. The third collision for the horizontal breather
occurs at $t_3\simeq 68$ and $t_3-t_2\simeq 9$ and so on.

(ii) Second, the amplitude of the horizonal breather decays with
time. The soliton and the antisoliton in the vertical breather can
not change their $\pm 2\pi$ amplitudes, instead they slow down and
eventually form a static pair.

(iii) Travelling waves are emanating from the breather after the
collision.

\begin{figure*}[t]
\includegraphics[width=6.5 in]{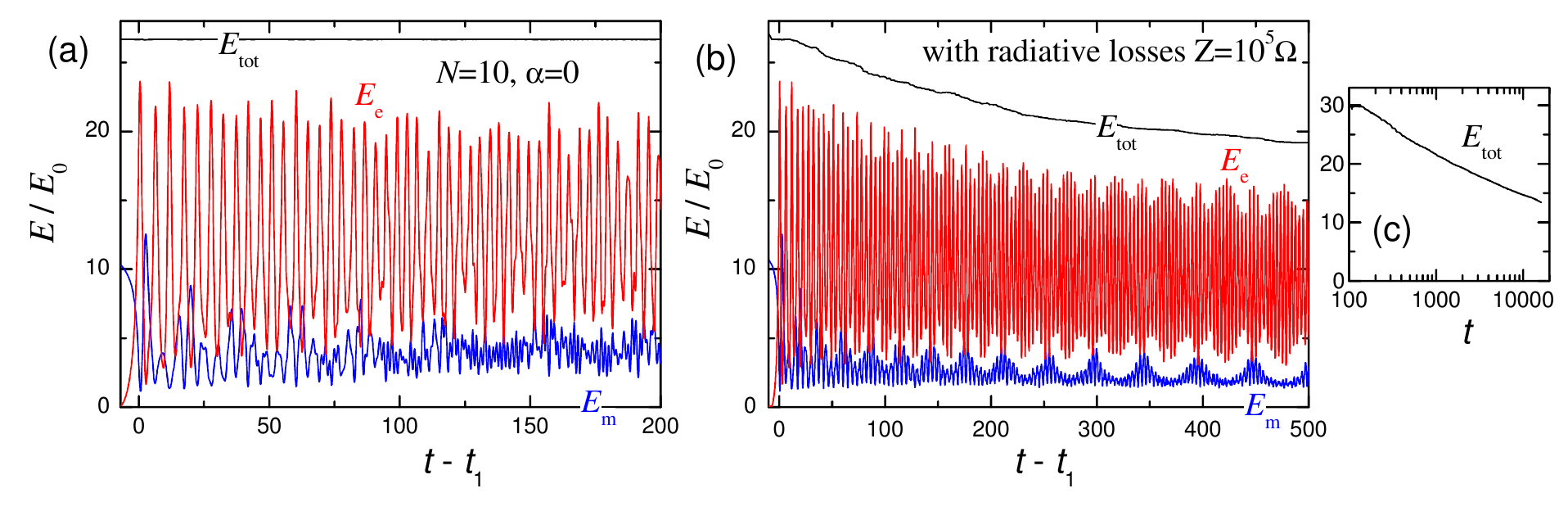}
\label{Fig3} \caption{ (Color online). Time dependence of the
total $E_{tot}$, electric $E_e$ and magnetic $E_m$ energies for a
horizontal breather in junction $i=5$ of the unperturbed $N=10$
CSGE system. (a) Without radiative losses at the edges $Z=\infty$,
(b) with radiative losses $Z=10^5~\Omega$. Panel (c) shows a
long-time evolution of $E_{tot}(t)$ in the presence of radiative
losses. A peculiar logarithmic time decay is seen.}
\end{figure*}

Figure 2 shows time dependence of $\varphi_{2}$ for the horizonal
breather from Fig. 1 (a). For simplification the phase $\varphi_1$
(thick lines) is shown only at the moments of collisions, marked
by arrows. It is clearly seen that waves are emitted from the
breather upon the first soliton-antisoliton collision $t_1\simeq
47$. Two wave fronts can be distinguished: the faster (marked by
the lower dashed blue lines) propagate with a constant speed $c_1
= 1.414$, and the slower (marked by the lower dotted green lines)
with the speed $c_2 = 0.8165$ in agreement with Eq. (\ref{c_n}). A
comparison of $\varphi_{1,2}$ clearly demonstrates that the faster
front has the in-phase $\varphi_1=\varphi_2$ and the slower the
out-of-phase eigenmode symmetry. At the second collision
$t_2\simeq 59$, two new wave fronts are emitted, marked by the
upper dashed and dotted lines, originated at the second collision
point ($x=0$, $t=59$). Every subsequent collision leads to a
similar emission. We emphasize that this happens {\it in the
absence of dissipation} $\alpha =0$. Therefore, the breather is
decaying in the unperturbed CSGE entirely due to radiative losses.

Figure 3 (a) shows time dependence (counted from the first
collision $t_1$) of the total energy $E_{tot}$ (top black line),
the electric energy $E_e$, given by the third term in Eq.
(\ref{Energy})(middle red line), and magnetic energy $E_m$, given
by the first term in Eq. (\ref{Energy}) (bottom blue line) for the
case of a horizontal breather in the middle $i=5$ of $N=10$
coupled junctions with the length $L=100$. Calculations are made
for the unperturbed CSGE $\alpha=\gamma=0$ and without radiation
emission at the edges $Z=\infty$. It is seen that $E_{tot}$ is
conserved because there are no dissipative or radiative losses.
Maxima in $E_e$ and minima in $E_m$ occur upon soliton-antisoliton
collisions. It is seen that the period of collisions is decreasing
with time and the magnetic energy $E_m$, related to the breather
amplitude, is rapidly decreasing after the first collision. This
is similar to the $N=2$ case shown in Fig. 1 (a). At $t-t_1
> L/ c_n$ the emitted waves from the breather reflect back from
the edges and come back to the breather. This leads to a very
complicated phase pattern consisting of a breather and bouncing
waves from all $N=10$ eigenmodes.

In order to avoid a complication associated with the reflection
and bouncing of the emitted waves, we made simulations for the
same parameters with a finite radiative impedance $Z$
\cite{TheoryFiske}. In this case the waves partly transmit through
the edges and leave the system. This leads to a decay of
propagating waves, except for the destructively interfering
out-of-phase mode $n=N$, which can not be emitted (see Ref.
\cite{TheoryFiske} for a discussion of emission efficiency of
different eigenmodes).

Fig. 3 (b) presents the results of simulations for the same case
as in (a) but with the finite radiative impedance. It is seen that
initially the total energy is conserved until $t-t_1 =
L/2c_1\simeq 10$ when the fastest in-phase ($n=1$) wave front
reaches the edges. After that the energy starts to decay due to
radiative losses through the edges. At large times the decay of
the breather energy $E_{tot}(t)$ slows down. Simultaneously, long
period beatings in $E_m$ due to slow flexural oscillations of the
breather become obvious.

Do the soliton and the antisoliton completely annihilate upon
direct collision in the unperturbed CSGE, or do they eventually
form a stable non-decaying breather? To answer this question we
performed long-time calculations. In order to avoid possible
radiative losses at the edges from the tail of the breather
itself, we studied even longer systems. Fig. 3 (c) shows such
simulations for $L=300$ and the rest of parameters the same as in
panel (b). A peculiar logarithmic time decay is clearly seen:
\begin{equation}\label{LogDecay}
E_{tot}(t) \simeq E_{tot}(0) - \beta \log (t/t_1),
\end{equation}
where $\beta$ depends on the strength of radiative losses at the
edges. Thus, unlike in the unperturbed sine-Gordon equation, in
the unperturbed coupled sine-Gordon equation the
soliton-antisoliton pair does annihilate upon the direct collision
even in the absence of dissipation, but the annihilation takes an
exponentially long time. Therefore, for all practical cases the
breather would appear stable at the time of the experiment, just
like the circulating current in type-II superconductors
\cite{Tinkham}.

The indirect soliton-antisoliton collision, shown in Fig.
\ref{Fig1} (b), does not lead to annihilation, but to formation of
a static bound pair. The vertical breather, produced upon the
indirect collision decays in a similar way as the horizontal
breather discussed above. The total radiative losses upon the
indirect collision is the difference between the energy of two
isolated solitons and the static vertical soliton-antisoliton
pair. For the double junction, shown in Fig. 1 (b), those are
given by Eqs. (\ref{Esol_N2},\ref{E_vertical2}):
\begin{equation}\label{Delta_E_N2}
\Delta E(N=2) = 8\left[(1-S)^{-1/2}-(1+S)^{-1/2}\right].
\end{equation}
For $S=0.5$, as in Fig. 1 (b), about $27\%$ of the initial energy
is lost into radiation.

\subsection{Soliton-antisoliton dynamics in the perturbed CSGE}

Addition of viscous damping $\alpha >0$ perturbation term leads to
decay of the breather both in the single SG and in the CSGE. As
seen from Fig. 3, reduction of the amplitude of breather
oscillations is accompanied by the increment of oscillation
frequency. Figure 4 shows the amplitude of the breather as a
function of breather frequency for the SG ($N=1$) and the
horizontal breather in the CSGE ($N=2$ and 10) with different
damping $\alpha$. It is seen that in all cases the
$\varphi_{Br}(\omega_{Br})$ dependence follows well the
theoretical expression Eq. (\ref{Breather}), shown by the dashed
line. The scattering of points for the CSGE case is due to
presence of the strong radiative field, which complicates the
determination of the breather frequency and amplitude, see Fig. 3
(a).

\begin{figure}[t]
\includegraphics[width=3.0 in]{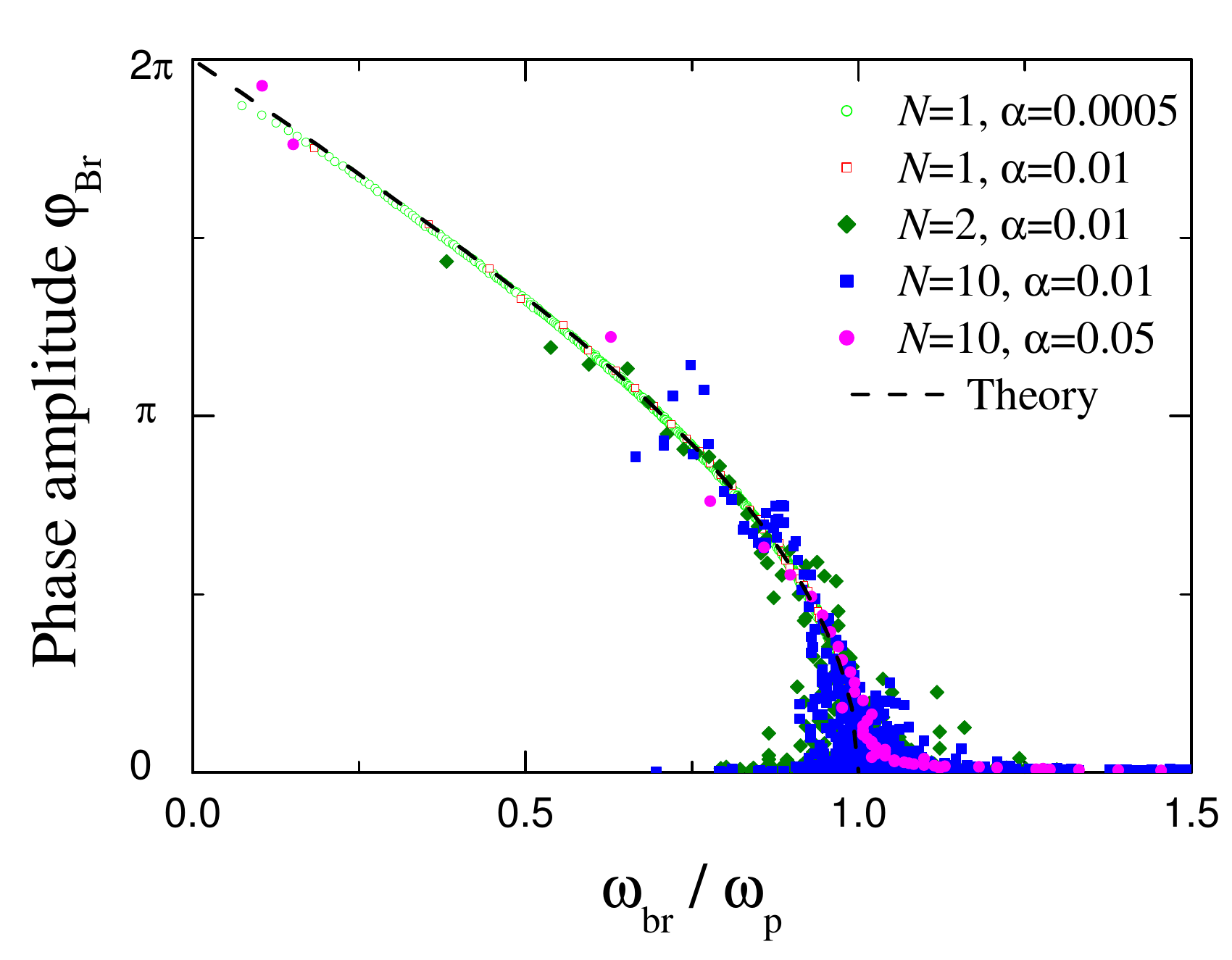}
\label{Fig4} \caption{ (Color online). Phase amplitude of the
horizontal breather as a function of the breather frequency for
perturbed SG ($N=1$) and CSGE with $N=2$ and 10 for different
values of the damping $\alpha$ and for $\gamma=0$. The dashed line
represents the breather solution Eq. (\ref{Breather}) for the SG
equation. }
\end{figure}

\subsection{Collision of a driven soliton with the edge}

So far we considered the case without driving force $\gamma=0$. As
shown above, in the CSGE soliton and antisoliton always annihilate
due to either radiative or dissipative losses. The driving force
$\gamma \neq 0$ replenishes the energy lost in the collision and
may lead to survival of the solitons after the collision.

In a finite system, a moving soliton will inevitably collide with
the edges. For the zero-field boundary condition the collision of
the soliton with the edge is equivalent to a collision with an
image antisoliton \cite{Scott}. After the collision the image
soliton continues the motion, i.e. the soliton is reflected as an
antisoliton at the edge \cite{Scott}. The shuttling
soliton-antisoliton motion leads to appearance of zero-field steps
(ZFS) in current-voltage ($I$-$V$) characteristics of Josephson
junctions
\cite{PedersenAnnih,PedersenSol,PedersenZFS,Binslev,Chang}. In
this case, current is the dc-driving force $I=\gamma$ and dc
voltage is time-average of the velocity.

\begin{figure}[t]
\includegraphics[width=3.0 in]{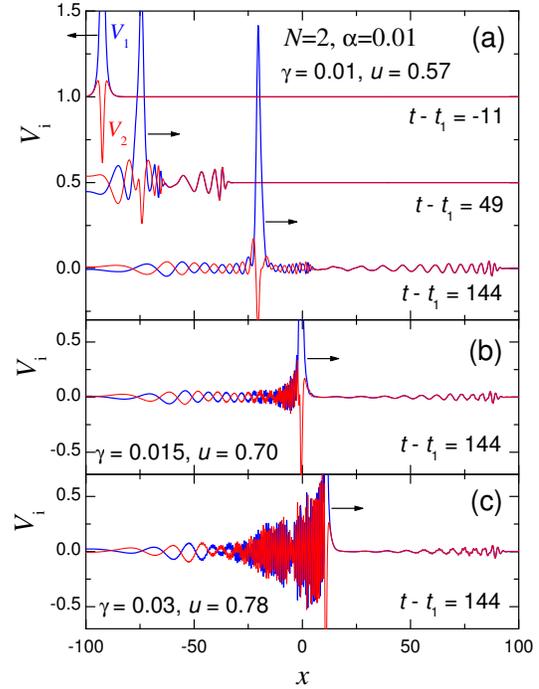}
\label{Fig5} \caption{ (Color online). Instantaneous voltage
(velocity) $V(x)=\partial\varphi_{1,2}/\partial t$ profiles for a
driven soliton motion in a dissipative $\alpha=0.01$ CSGE with
$N=2$ for increasing driving currents (forces) (a) $\gamma=0.01$,
(b) $\gamma=0.015$ and (c) $\gamma=0.03$. Time is counted with
respect to the first collision $t_1$ with the left edge of the
system. It is seen that the fast in-phase and the slow
out-of-phase waves are emitted upon the collision $t=t_1$. In
panel (b) the soliton velocity $u$ is close to the velocity of the
out-of-phase wave, and the corresponding front is no longer seen
ahead of the soliton. Further increase of $u$ in panel (c) leads
to profound Cherenkov-type radiation behind the soliton. }
\end{figure}

Figure 5 shows snapshots of voltage profiles $V_i=\dot \varphi_i$
for a single soliton in the junction $i=1$ of a double junction
structure $N=2$ for different driving terms $\gamma$ and for
$\alpha=0.01$ and $L=200$. The time is counted relative to the
first collision $t_1$ with the left edge. Panel (a) corresponds to
a small driving force $\gamma=0.01$ and a slow, subluminal soliton
motion $u=0.57 <c_N$. Before the collision, $t-t_1=-11$, the
soliton was moving to the left. After collision it is reflected as
an antisoliton moving to the right. Simultaneously, emission of
plasma waves from both the in-phase and the out-of-phase
eigenmodes takes place, similar to the unperturbed case in Figs. 1
(a) and 2. Panels (b) and (c) show snapshots at larger driving
forces and soliton velocities larger than the out-of-phase plasma
wave speed $c_N$. Such superluminal soliton motion is accompanied
by Cherenkov-type radiation behind the soliton
\cite{Modes,Fluxon,Cherenkov,Goldobin}. A comparison of snapshots
at $t-t_1=144$ in panels (a-c) indicates that the in-phase
radiation front from the collision event is similar for all shown
soliton velocities, but the out-of-phase front is not visible
ahead of the soliton when the soliton is moving faster than the
out-of-phase plasma wave \cite{Note2}. From this it is also clear
that it is the edge, rather than the moving soliton, that emanates
the waves.

\begin{figure*}[t]
\includegraphics[width=7.0in]{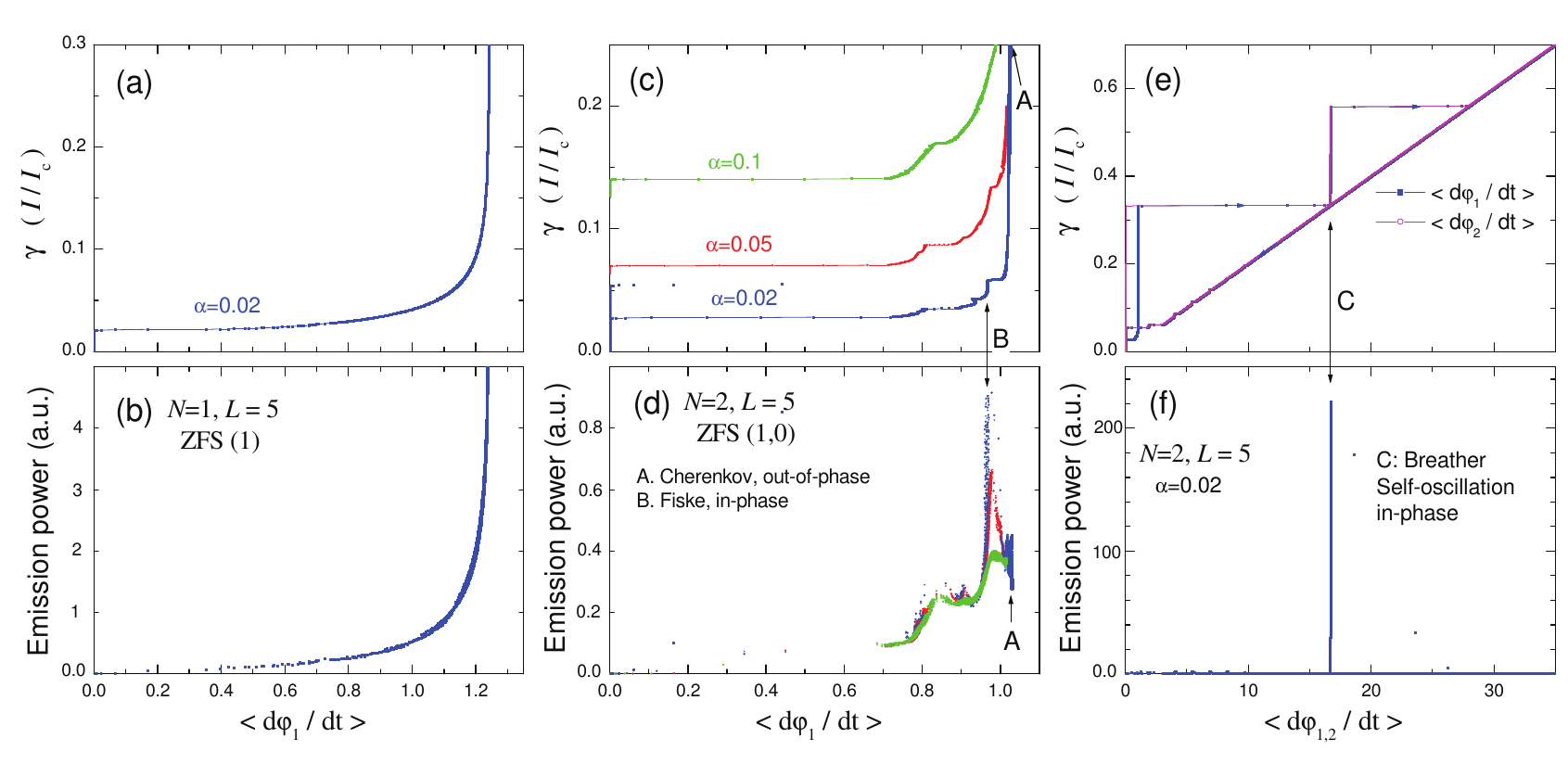}
\label{Fig6} \caption{ (Color online). Current (driving force) -
Voltage (velocity) characteristics of a shuttling single soliton
in a moderate size systems $L=5$ for (a) a single junction (SG,
$N=1$) and (c) a double junction stack (CSGE, $N=2$), appearance
of a fine structure of the zero-field step due to interference
with emitted plasma waves is clearly seen. Panel (e) shows the
continuation of $I-V$ characteristics at large bias for $N=2$ and
$\alpha=0.02$. Panels (b), (d) and (f) show the corresponding
emission powers. It is seen that in the double junction system the
emission at the velocity matching part of the zero-field step
(point A in (c,d)) is at minimum, unlike the single junction case
(a,b), and the maximum emission occurs at point B, corresponding
to the in-phase geometrical resonance and the Fiske step in the
$I-V$. Above the ZFS, the system switches to another strongly
emitting resonance (point C in (e,f)), which represents a
breather-type self-oscillation. }
\end{figure*}

\subsection{Soliton resonances in finite-size systems}

A shuttling soliton in a finite-size system will periodically
excite travelling waves at the edges. The emanated waves also
propagate along the chains and reflect back at the edges. The
shuttling soliton will interact with bouncing waves. Resonance
will appear if bouncing waves are in-phase with the soliton at the
edges \cite{Golubov}. In Josephson junctions this leads to
appearance of fine structure of Zero Field Steps in $I$-$V$
characteristics \cite{Chang,PedersenSol}.

Figure 6 (c) shows calculated dc current-voltage $\gamma$ -
$<\partial\varphi_1/\partial t>$ ($<>$ indicate averaging in time)
characteristics for a moderately short $L=5$ double junction $N=2$
structure for different damping parameters $\alpha$. Calculations
are made for the ZFS mode (1,0), i.e. for a single soliton
shuttling in the junction $i=1$. In the second junction
$<\partial\varphi_2/\partial t>=0$. The dc ZFS voltage is
\begin{equation}\label{V_ZFS}
V_{ZFS}(1,0)=\langle\frac{\partial \varphi_1}{\partial t}\rangle=
\frac{2\pi u}{L}.
\end{equation}
A strong almost vertical velocity-matching soliton step, marked as
point A in Fig. 6 (c), occurs when the soliton velocity approaches
the slowest out-of phase eigenmode velocity $u\rightarrow c_2$.
According to Eqs. (\ref{c_n}) and (\ref{V_ZFS}), for $S=0.5$,
$N=2$ this occurs at $<\partial\varphi_1/\partial t>=1.026$. The
rapid increase of the driving force $\gamma$ at $u\rightarrow c_2$
is caused both by a partial Lorentz contraction of the $F_2$
component of the composite soliton \cite{Modes} and by a rapid
enhancement of dissipation due to Cherenkov radiation, see Fig. 5
(b,c).

Fine structure is seen at the ZFS below the velocity matching step
due to resonances between the shuttling soliton and bouncing waves
emanating upon every collision of the soliton with edges. In
principle, the soliton can interfere and form resonances with any
type of periodically emitted waves. Those can be waves emitted by
the fluxon upon passing a defect \cite{Golubov} or Cherenkov-type
emission \cite{Goldobin}. However, the resonances seen in Fig. 6
(c) are different from the previously discussed types. Indeed, we
consider an ideal system without defects and Cherenkov emission
does not take place at the corresponding soliton velocities, as
demonstrated in Fig. 5 (a).

The observed fine structure of ZFS is due to inelastic nature of
soliton collision in the CSGE, even in the absence of
perturbations, as shown in Figs. 1 and 2. This is specific for the
CSGE and is not present in the unperturbed SG equation, in which
the soliton collision is always elastic. Even though, some
radiation appears in the SG in the presence of dissipation $\gamma
\ne 0$, the effect is very small \cite{Kivsh_Malomed_RMP}. For
comparison, in Fig. 6 (a) we show ZFS in a single junction case
($N=1$), calculated for the perturbed SG with the same parameters
as in Fig. 6 (c). The velocity-matching step at $u\rightarrow 1$
and $<\partial \varphi / \partial t> = 2\pi/L \simeq 1.26$ is
clearly seen. Unlike the CSGE case, Fig. 6 (c), it is entirely due
to relativistic Lorentz contraction of the soliton, Eq.
(\ref{Soliton}). Because in this case soliton - image antisoliton
collision at the edges is (almost) elastic, the fine structure is
not visible (a closer inspection reveals the presence of tiny
wiggles at the ZFS).

\subsection{Radiation emission}

Coupled systems are interesting from the point of view of
achieving coherent superradiant emission. In particular, THz
emission from stacked intrinsic Josephson junctions in cuprate
superconductors at zero applied magnetic field is being actively
discussed \cite{Ozyuzer,Hu,TheoryFiske,Breather}.

To estimate the emission from the stack at the ZFS, we employed
the dynamic radiative boundary conditions, as in Ref.
\cite{TheoryFiske}. The effective radiative impedance was very
large 
so that radiative losses do not affect soliton dynamics. The
emission power from the left edge of the double junction stack is
shown in Fig. 6 (d). Noticeably, the emission at the
velocity-matching step A is at minimum, despite a large
dissipation power $P=IV$, because at point A the soliton is
resonating with the (Cherenkov) out-of-phase eigenmode $n=N=2$,
see Fig. 5 (c). Even though the oscillation amplitude in each
junction is very large, destructive interference from the two
junctions prevents emission \cite{TheoryFiske}. This is
qualitatively different from the single junction $N=1$ case, shown
in Fig. 6 (b), in which the maximum emission occurs at the
velocity matching step and the emission power is correlated with
the total power $IV$.

From Fig. 6 (d) it is seen that the main emission occurs at the
lower resonance B, which corresponds to the voltage of the
in-phase cavity (Fiske) mode $(m,n)=(2,1)$ in the stack
\cite{KatterweFiske,TheoryFiske}:
\begin{equation}\label{V_fiske}
V_{m,n}=m\frac{\pi c_n}{2L}.
\end{equation}
At this point the shuttling soliton excites the in-phase standing
wave in both junctions, which leads to constructive interference
and to significant superradiant emission outside the stack
\cite{TheoryFiske}. Note that the emission power at the resonance
B is rapidly increasing with decreasing $\alpha$, unlike for the
rest of the ZFS. This is a clear indication that the geometrical
resonance is indeed taking place at point $B$, because the
emission at the Fiske step depends on the quality factor of the
geometrical resonance and increases with decreasing $\alpha$
\cite{TheoryFiske}.

The appearance of the Fiske step at the ZFS provides a clear
evidence that the shuttling soliton in the CSGE can indeed
strongly interact with cavity modes and travelling waves due to
strong emission upon soliton - image antisoliton collisions at the
edges. In a similar manner, the soliton also interacts with
Josephson oscillations when one or several junctions are in the
running (McCumber) state with $<\partial \varphi_i / \partial t>
\simeq \gamma / \alpha$. This leads to a large variety of resonant
states \cite{KleinerAntiFlux} and may lead to self-oscillation
phenomena at geometrical resonance conditions \cite{Breather}.

Figure 6 (e) shows the continuation of the $I$-$V$ for the same
$N=2$ double junction with $\alpha=0.02$ up to higher bias
current. It is seen that at $\gamma > 0.33$ the system switches
from the ZFS (1,0) to another strong resonance C before it goes
into the Ohmic (free running) state. At this resonance both
junctions have the same voltage and are synchronized in-phase.
This leads to a large emission, as shown in panel (f). Such
resonances were discussed in Refs.
\cite{KleinerAntiFlux,Hu,Breather}. They may combine $2\pi$
soliton kinks with a similarly large amplitude waves, which may be
difficult to disentangle by just looking at the shapes of phase
profiles $\varphi_i(x)$. However, we observed that high-order ZFS
can be clearly distinguished from geometrical resonances by
comparing the emission frequency: ZFS emit at the sub-harmonics of
the Josephson frequency \cite{PedersenZFS} or even at
non-Josephson frequency \cite{Binslev}, while self-oscillations at
geometrical resonances emit at the harmonics of the Josephson
frequency \cite{Breather}.

To conclude, we have studied soliton-antisoliton collisions in the
coupled sine-Gordon equation. It was shown that in contrast to the
sine-Gordon equation, a soliton-antisoliton pair annihilates in
the CSGE even in the absence of perturbations. The annihilation
occurs via a logarithmic-in-time decay of a breather caused by
emission of plasma waves. In a dissipative, dc-driven case, a
similar phenomenon leads to a strong coupling between the coupled
soliton-antisoliton pairs, breathers, and propagating waves, which
may lead to self-oscillations at the geometrical resonance
conditions. This phenomenon may be useful for achieving
superradiant emission from coupled oscillators.

\end{document}